# User-Centered Design of Socially Assistive Robotic Combined with Non-Immersive Virtual Reality-based Dyadic Activities for Older Adults Residing in Long Term Care Facilities

Ritam Ghosh[1*], Nibraas Khan[2], Miroslava Migovich[3], Judith A. Tate[4], Cathy Maxwell[5], Emily Latshaw[4], Paul Newhouse[6,7], Douglas W. Scharre[8], Alai Tan[4], Kelley Colopietro[3], Lorraine C. Mion[4], Nilanjan Sarkar[1,2,3]

[1*]Department of Electrical and Computer Engineering, Vanderbilt University, 400 24th Ave S, Nashville, 37212, Tennessee, United States.

[2]Department of Computer Science, Vanderbilt University, 1400 18th Avenue South, Nashville, 37212, Tennessee, United States.

[3]Department of Mechanical Engineering, Vanderbilt University, 2400 Highland Avenue, Nashville, 37212, Tennessee, United States.

[4]College of Nursing, The Ohio State University, 1585 Neil Avenue, Columbus, 43210, Ohio, United States.

[5]School of Nursing, Vanderbilt University, 461 21st Avenue South, Nashville, 37240, Tennessee, United States.

[6]Department of Psychiatry and Behavioral Sciences, Vanderbilt University Medical Center, 1211 Medical Center Dr, Nashville, 37232, Tennessee, United States.

[7]Tennessee Valley Geriatric Research, Education, and Clinical Center (GRECC), U.S. Department of Veterans Affairs, 1310 24th Avenue South, Nashville, 37212, Tennessee, United States.

[8]Department of Neurology, The Ohio State University Wexner Medical Center, 395 W. 12th Avenue, Columbus, 43210, Ohio, United States.

**\*Corresponding author(s). E-mail(s): ritam.ghosh@Vanderbilt.Edu**

Acknowledgments

Research reported in this publication was supported by the National Institute on Aging of the National Institutes of Health under award number R01AG062685.

Abstract (150-250 words)

Apathy impairs the quality of life for older adults and their care providers. While few pharmacological remedies exist, current non-pharmacologic approaches are resource intensive. To address these concerns, this study utilizes a user-centered design (UCD) process to develop and test a set of dyadic activities that provide physical, cognitive, and social stimuli to older adults residing in long-term care (LTC) communities. Within the design, a novel framework that combines socially assistive robots and non-immersive virtual reality (SAR-VR) emphasizing human-robot interaction (HRI) and human-computer interaction (HCI) is utilized with the roles of the robots being coach and entertainer. An interdisciplinary team of engineers, nurses, and physicians collaborated with an advisory panel comprising LTC activity coordinators, staff, and residents to prototype the activities. The study resulted in four virtual activities: three with the humanoid robot, Nao, and one with the animal robot, Aibo. Fourteen participants tested the acceptability of the different components of the system and provided feedback at different stages of development. Participant approval increased significantly over successive iterations of the system highlighting the importance of stakeholder feedback. Five LTC staff members successfully set up the system with minimal help from the researchers, demonstrating the usability of the system for caregivers. Rationale for activity selection, design changes, and both quantitative and qualitative results on the acceptability and usability of the system have been presented. The paper discusses the challenges encountered in developing activities for older adults in LTCs and underscores the necessity of the UCD process to address them.

# 1 Introduction

In 2020, the population of older adults (age 65 or older) in the US was 55.6 million, representing 17% of the total population and expected to grow to represent 22% by 2040 [1]. Many older adults have multiple chronic health conditions that impact physical and cognitive function, resulting in a need for assistance with activities of daily living (ADL) and instrumental activities of daily living (IADL). Based on the severity of their functional impairments and available resources, older adults may require residence in long-term care settings (LTCs) to receive the type, frequency, and intensity of services necessary to maintain function or delay further decline. Indeed, according to the US Centers for Medicare and Medicaid, 63% of LTC residents have four or more ADL impairments and 75% have moderate to severe cognitive impairment [2].

In the U.S., 1.3 million older adults reside in nursing homes and another 918,000 reside in assisted living facilities [3], representing 3.8% of the older adult population. However, the lifetime probability of requiring LTC services is 70% for those who survive to age 65 and 15% will spend two or more years in a nursing home [4].

The presence of *apathy* complicates the care delivery to these vulnerable older LTC residents. Apathy is a condition that results in a lack of initiative, loss of interest in daily activities, and reluctance towards social interactions [5]. Older adults who suffer from apathy due to neurodegenerative disease or depression often remain disengaged with their surroundings and may require significant prompting and encouragement to perform basic activities, such as speaking, eating, bathing, or walking. Apathy is common among LTC residents, occurring in up to 82% of those with dementia and up to a third among those with depression [6, 7, 8]. Apathy has significant deleterious consequences including accelerated cognitive decline, further functional deficits, diminished quality of life, and increased rates of mortality [9]. Not only does apathy negatively impact the older adult, but it causes significant burden and stress on family and staff [5, 6].

Despite the scope and significance of apathy, few pharmacologic options are available [10] and so managing apathy requires effective nonpharmacologic strategies. Multimodal activities that combine motor-based and cognitive aspects with social, i.e., human-to-human interaction (HHI), appear to be the most effective for enhancing engagement and reducing apathy among older LTC adults [11, 12, 13, 14, 15].

Unfortunately, designing and delivering multimodal activities are resource intensive but many LTCs have inadequate staffing, either in labor quantity or skill [16, 17]. Thus, technology has been suggested to complement existing staff and facilitate the delivery of care to older adults with varying conditions [18, 19, 20]. Although the majority of technological interventions in LTCs have focused on the electronic health record, telehealth, and wearables, there is a growing interest in the use of non-immersive or immersive virtual reality (VR) [21, 22, 23] or socially assistive robots (SAR) [24, 25] to enhance function or quality of life. Our work focuses on the combination of SAR with non-immersive VR, hereinafter referred to as SAR-VR.

Both SARs and VR have shown promising results for older adults [21, 26, 24] and each has its strengths and limitations. SARs have the distinct advantage of physical presence, the ability to emulate body language, and use of non-verbal communication gestures. Although on-screen avatars are capable of gestures, people are more likely to trust physical robots over screen-based avatars [27]. However, the readily available commercial SARs have limited payload capability to manipulate physical objects and custom SARs are expensive and often designed for a specific application. Non-immersive VR-based systems allow for the creation of various physical scenarios displayed on a standard computer monitor that provides a potentially greater range of activities compared to SARs. VR activities can also be easily adjusted to accommodate an individual's physical capabilities (e.g., impaired mobility, reduced strength, diminished range of motion). However, non-immersive VR limits the interaction to use of a mouse, joystick, or remote control, which may be difficult for older adults with reduced motor dexterity and proprioception in hands and wrists [28, 29, 30].

Recent work has demonstrated the successful combination of assistive robotics with non-immersive VR for other populations including stroke rehabilitation [31, 32], rehabilitation for children with motor skill impairment [33], and conducting video lectures [34]. In our earlier pilot work [35], we successfully combined an SAR with a non-immersive VR-based activity using gesture control for engaging pairs of older adults residing in the community and subsequently in LTCs. The activity was designed based on input from geriatric experts and refined based on older adults' usability testing.

Our long-term goal is to utilize SAR-VR systems and dyadic activities to address apathy among older LTC residents. As an initial step, our aim was to establish a library of SAR-VR dyadic activities that combined cognitive, physical, and social domains and encouraged human-to-human (HHI) engagement during the activities. Given the range of physical and cognitive impairments in this population, involving older LTC adults in the development of interactive health technologies will help to enhance the functionality, usability, and likelihood in promoting the intended health outcome [36, 37]. However, older LTC adults are often absent during the design of interactive health technologies [38].

To enhance the likelihood that the designed SAR-VR system and dyadic activities could effectively engage older LTC adults with varying levels of physical and cognitive function, we utilized a user-centered design approach (UCD) [36]. UCD is supported by various tools and methods [39], but little information is available about the practical implementation of UCD methods with older LTC adults. Challenges to UCD implementation with this population include not only the characteristics of the older adults, but also organizational and staff characteristics. To address gaps in knowledge and facilitate future work with this vulnerable population, we present ours and stakeholders' experiences in conducting UCD, characterize the challenges encountered in developing the SAR-VR system and dyadic activities, and present the quantitative and qualitative results of the UCD. At the beginning of this study, we hypothesized that the successive iterative designs during the UCD process will improve the user's comfort and confidence level with the different components of the system, and the results presented in section 7.1 validate that hypothesis.

## 2　UCD METHODS

### 2.1　Study Design

We implemented a multi-step, user-centered design (UCD) process to design and evaluate SAR-VR dyadic multimodal activities that encourages human-human interaction (HHI) through human-computer interaction (HCI) and human-robot interaction (HRI). Three overarching UCD principles guided the research process throughout: focus on users and tasks, measure usability empirically, and design and test usability iteratively [40, 41]. Fig. 1 displays the UCD timeline.

#### 2.1.1　Phase 1: Prototype Development

The first phase focused on prototype development of dyadic SAR-VR activities. The initial focus was on problem analysis, i.e., to specify the context of use of dyadic SAR-VR activities with frail older adults. Our interdisciplinary team brought expertise to the area of inquiry: engineers from various disciplines (e.g., mechanical engineering, electrical engineering, and computer science), nurses, and physicians specializing in geriatrics and long-term care. We searched PubMed to a) identify aspects of activities that effectively engage older LTC adults, b) SAR characteristics acceptable to older adults, and c) non-immersive VR factors suitable to older adults. We assembled an interdisciplinary advisory panel consisting of occupational therapists specializing in geriatrics, LTC activity directors, LTC nurses, and LTC managers who were experienced in working with older adults.

Based on the literature review and conversations with the advisory panel and older adults, inspiration for activities was based on real-life activities the older adults performed at their LTCs and hobbies they enjoyed. We conducted multiple design iterations and internal testing of prototype activities in the VU Engineering Laboratory. We presented ongoing work over several meetings with the advisory panel. Two LTC residents reviewed initial prototype activities via teleconferencing due to COVID-19 isolation protocols. We refined the prototype based on their suggestions for minor cosmetic changes.

#### 2.1.2　Phase 2: Determining Usability and Measures

Usability requirements of the SAR-VR dyadic activities and system were based on the intended end-users: LTC older adults and LTC staff. Activity requirements were the flexibility to accommodate older adults with varying levels of physical and cognitive abilities, require some degree of collaboration, and include sufficient variation to minimize monotony. The system must have a user-friendly graphical user interface so that LTC staff can easily operate it and for older adults to manipulate.

Usability assessment reports, using a pre-determined format, were generated at each testing session, and reviewed by the research team in light of the above requirements. Data were gathered via observation, video recordings, physiological readings, and interviews after each session.

### 2.1.3 Iterative Testing and Evaluation

Two successive rounds of field testing were conducted in separate LTC sites using the intended deployment setup to enhance discovery of potential challenges. During each round of field testing, multiple sessions were conducted with testing and analysis that required looping back to earlier stages of the prototype development and defined requirements for cyclical refinement and improvement of both the activity and the system. The design iterations continued until users deemed the dyadic activities and SAR-VR system met the requirements as represented by their approval ratings and qualitative assessment presented in section 7.

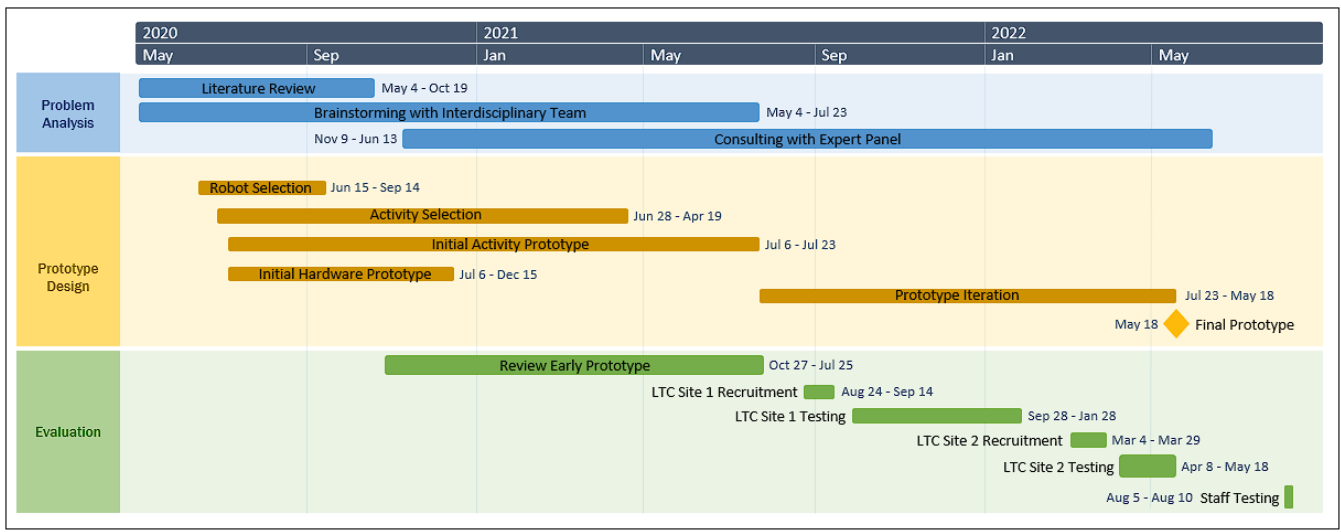

Fig. 1: Timeline of User Centered Design Process

## 2.2 Compliance with Ethical Standards

Approval for this study was obtained from the Institutional Review Boards of the research team: The Ohio State University and Vanderbilt University. The procedures used in this study adhere to the tenets of the Declaration of Helsinki [42] and regulations of the Institutional Review Boards. Written informed consent was obtained from all individuals participating in the study.

## 3 SAR-VR System Architecture

The long-term goal of the SAR-VR system and dyadic activities is to increase social engagement (i.e., HHI) among older LTC adults with apathy through collaborative activities mediated by Human-Robot Interaction (HRI) and Human-Computer Interaction (HCI). To achieve this, we designed a system architecture that encourages two participants to perform an activity together in a non-immersive virtual environment. Activities were designed specifically for use with a humanoid SAR and a dog SAR. The humanoid SAR acted as coach and cheerleader, while a computer-based avatar acted as the coach and cheerleader for activities involving the dog SAR. The architecture consists of three major modules: an HCI, an HRI, and an HHI module, as shown in Fig. 2.

The *HCI module* is responsible for providing the participants with a virtual environment where they can perform the activities. It consists of a Windows computer running the VR environment designed using the Unity game engine (https://unity.com/) and a set of two 'wands', one for each participant, which are custom-designed controllers to interact with the virtual activities [41, 43]. The wands function similar to Wii remote controllers; they contain inertial measurement units (IMU), which tracks the participant's hand motion and control virtual objects on the screen, a vibration motor to provide tactile feedback and two buttons to select objects. They communicate with the computer wirelessly via a dongle using the ESP-NOW protocol [44]. A finite state machine controls the different states of the activities.

The HRI module is responsible for using SARs to demonstrate the virtual activities to the participants, provide corrective feedback when necessary, encourage and motivate participants to keep them engaged, and provide rewards in the form of celebratory dance, tricks etc. upon successful completion of the activities. The HRI module can use two SARs, a humanoid robot Nao (www.softbankrobotics.com), and a puppy robot Aibo (us.aibo.com). The HRI module communicates with Nao via a socket connection wirelessly through a router and with Aibo through an Application Programming Interface (API) key. In our earlier work [41, 43], we found that older adults varied in their responsiveness and enjoyment of the two types of SARs; using two different forms will allow us to examine the characteristics of older adults best served by SAR type for future applications. The HRI module enables communication between the HCI module and the SARs. Nao acts as a coach and guides the participants through interactive tutorials before each VR activity. Nao also provides encouragement and corrective feedback to keep the participants engaged throughout the VR activity. Aibo entertains the participants and incentivizes the completion of VR activities by performing tricks as rewards for activity completion. In the final iteration, for the activity with the dog robot, an on-screen avatar acts as a coach and delivers encouragement and feedback through the VR environment.

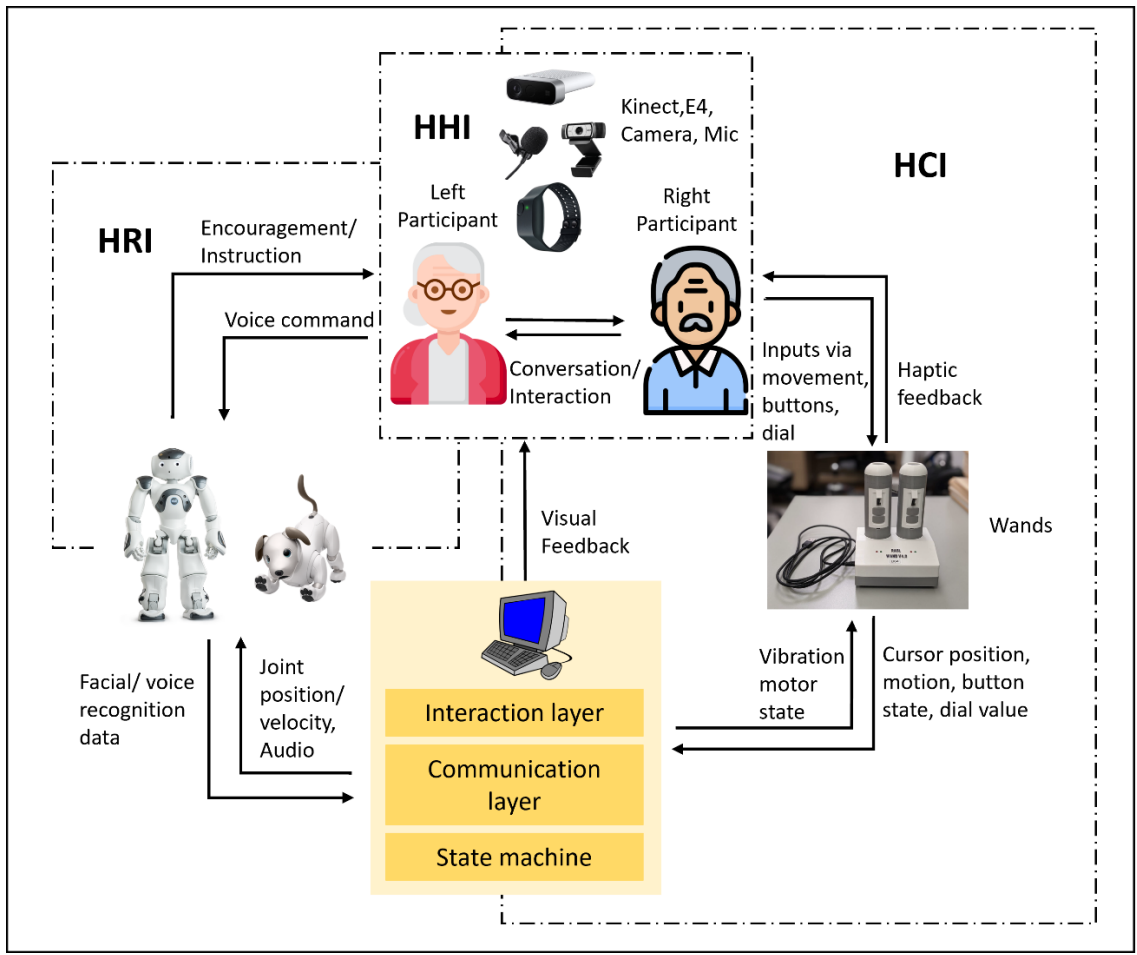

Fig. 2: SAR-VR Architecture

The HHI module was deliberately designed to require the participants to communicate and collaborate to complete the activities because existing literature found that HHI was particularly effective in reducing apathy [11, 13, 14]. HHI, in the form of non-verbal communication, is measured using participants' body position and orientation data collected using a Kinect sensor [45]; verbal communication is measured from video recordings of the sessions. Multimodal physiological data are collected using the Empatica E4 sensor (Empatica.com) for future studies that may provide an insight into the participants' physiological state during the performance of the activities. The audio, video, and data from all the sensors are synchronized using time stamps. A user-friendly menu system enables easy navigation of the system and selection of activities. Details of the system architecture can be found in Online Resource 1.

## 4    Early Prototype Design

The interdisciplinary team determined potential virtual activities that a) combined cognitive, physical and HHI components, b) could involve two older adults and encourage them to communicate and interact, and c) were suitable for LTC-residing older adults with varying levels of cognitive and physical capabilities. We selected Nao (www.softbankrobotics.com) as our humanoid SAR and Aibo by Sony (us.aibo.com) as the animal SAR based on prevalence of use in the literature, cost, and ease of use. These initial activity ideas were reviewed with our advisory panel and two LTC residents via teleconferencing (due to COVID-19 restrictions). Older adults (OAs) were also shown the two SARs; Nao introduced itself and danced while Aibo walked about and performed tricks. The residents provided valuable feedback on the activity ideas and indicated that they would like to interact with the robots.

### 4.1 Initial activity choice

Based on repeated brainstorming sessions, inspiration for activities was taken from real-life activities that the older adults performed at the LTCs and hobbies they enjoyed. We began development of four activities, three with Nao (music, fishing, painting) and one with Aibo (spelling) as shown in Fig. 3.

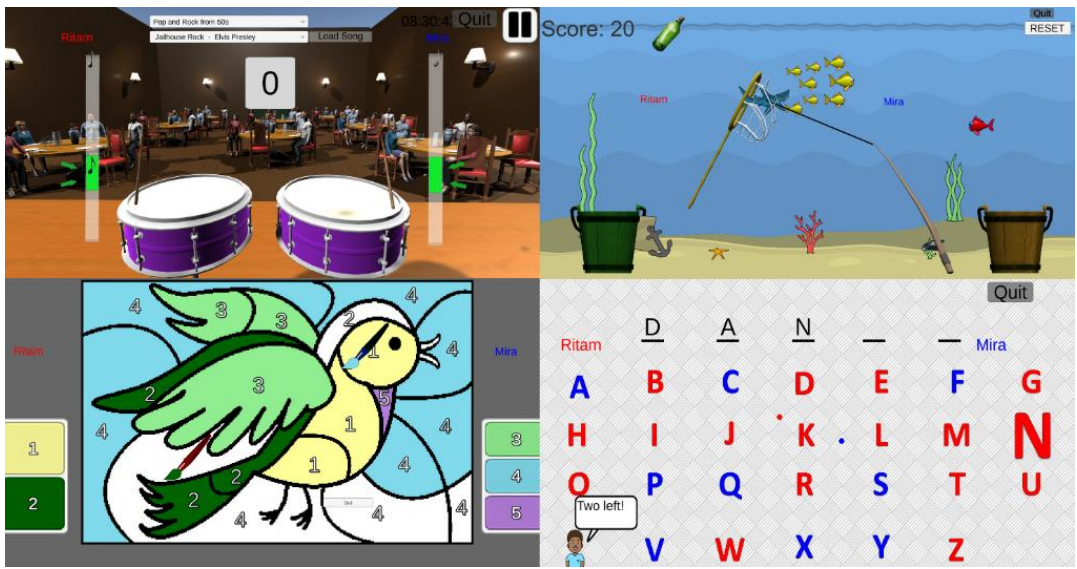

Fig. 3: Top left – Music Activity, top right – Fishing Activity, bottom left – Painting Activity, bottom right – Spelling Activity

#### 4.1.1 Rationale

Music can serve as an effective tool for reminiscence and autobiographical recall [46], improve cognitive function, and reduce depression [47]. The fishing and painting activities were chosen because many older adults enjoyed these activities in their youth. Research has also shown that creative engagement, such as art and music, can lead to improved mental well-being, increased focus, and a sense of calmness [48]. The virtual spelling activity was chosen since word games, such as scrabbles and crossword puzzles, are often enjoyed by older adults in LTCs and have shown promising results in the preservation of cognitive function [49, 50].

### 4.1.2 Activity design modifications

Each activity was structured to have varying physical, cognitive, and social components that required collaboration between the two participants (See Table 1). Each activity started with an interactive tutorial to familiarize participants with the objectives of the activity and how to operate the wand. Each activity had three to four levels of increasing difficulty to accommodate varying levels of cognitive abilities among the participants. All activities underwent numerous modifications as the field testing progressed based on a) enhancement of visual and audio characteristics and b) meeting our goal of encouraging human-human interaction through objective and subjective feedback. Major changes are highlighted in Table 1. A detailed description of the activities and changes made are available in Online Resource 2.

**Table 1.** Initial Prototype Activities and Refinements

|  | Music | Fishing | Painting | Spelling |
|---|---|---|---|---|
| **Description** | Drumming motion with wand to play virtual drums in sync with displayed musical notes. | Catch fish with virtual fishing rods using wands. Deposit captured fish into buckets. | Paint-by-numbers format. Each participant has assigned numbers with corresponding colors on palettes. | A word (dog command) provided to participants. Each participant is assigned color (red or blue) and must choose correct letters to complete the spelling. |
| **Cognitive skills** | Perception and attentional control. | Problem solving, attention, working memory. | Recognition and attention. | Recall |
| **Physical skills** | Gross motor movements. | Gross and fine motor movements. | Fine motor skills | Fine motor skills |
| **SAR** | Nao provides tutorials, encouragement, and reminders. | Nao provides tutorials, encouragement, and reminders. | Nao provides tutorials, encouragement, and reminders. | Nao provides tutorials and feedback. Aibo performs the word command. |
| **Field testing 1 changes** | Changed note generation from random to alternate. | Replaced left participant's rod with a net to collect the fish captured by right participant. | Replaced cursor with a brush. Brush tip changed to color selected. | Reminders added to cue participants when it was their turn to pick letter. |

|  | Music | Fishing | Painting | Spelling |
|---|---|---|---|---|
|  |  |  |  | Hint added to provide word if participants forgot. |
| **Field testing 2 changes** | Alternate notes are too predictable. Devised probability-based approach; initially participants had equal probability of receiving a note that decreased once they received a note. | At the highest level of difficulty, only one bucket active at a time. Required participants to pay attention to which bucket was activated. | Changed color palette from two column to single column palette. | Replaced Nao with onscreen avatar providing the same instructions and encouragement. Instructions also appear on screen as a speech bubble. |

## 5    Field Testing 1

### 5.1    Participants

Older adults were recruited from an LTC site within driving distance of Vanderbilt University. After confirming that the facility was interested in participating, the project and robots were presented to residents. Each activity was explained, and both robots were demonstrated. Following the presentation, interested residents were screened for study eligibility: age 65+ years, >3 months of residence at the facility, ability to hear, speak, and understand English, and the ability to sit comfortably and move both arms. Residents provided written informed consent.

Upon providing consent, participants completed a basic demographic questionnaire, the self-rated Apathy Evaluation Scale (AES-S), and the Self-Administered Gerocognitive Exam (SAGE). The AES-S [51] consists of 18 items that measures behavioral, cognitive, and emotional domains of apathy. Each item is rated on a 4-point scale ranging from 1 (not at all characteristic) to 4 (a lot characteristic) with a total score ranging from 18 (low apathy) to 72 (extreme apathy). The AES-S has good internal consistency (Cronbach's alpha = 0.86-0.94), retest reliability (r=0.89-0.94), and convergent and divergent validity. The SAGE [52] assesses cognition in 5 domains: language, reasoning/computation, visuospatial, executive, memory, and orientation. The SAGE has good interrater reliability (ICC = 0.96), specificity (88-95%), and sensitivity (62-95%). A score of 17-22 (maximum) suggests normal cognition, 15-16 suggests mild cognitive impairment (MCI), and <15 suggests dementia.

Six older adults initially consented to participate in the initial field study (Table 2). A 3-week delay occurred between consent and implementation due to a COVID outbreak in the facility. Subsequently, two participants chose not to continue with the study, having lost interest over time. The four remaining participants (mean age: 85 ±9.3, mean AES score: 29.25 ± 4.97 and mean SAGE score: 11.5 ±3.6) completed Field Testing 1.

**Table 2: Participant details for Field Testing 1**

| Participant ID | Age | Gender | Education | Technology use | AES-S score | SAGE score | Completed the study |
|---|---|---|---|---|---|---|---|
| A1001 | 89 | F | High School | Everyday | 25 | 11 | Y |
| A1002 | 90 | F | High School | Never use | 32 | 17 | Y |
| A1003 | 92 | M | Graduate degree | Everyday | 26 | 8 | N |
| A1004 | 92 | M | Bachelor's degree | Everyday | 24 | 11 | Y |
| A1005 | 69 | M | Graduate degree | Never use | 36 | 7 | Y |
| A1006 | 89 | F | High School | Never use | 25 | 5 | N |

### 5.2 Experimental Procedure

Fig. 4 shows the experimental setup, materials, and a session in progress. Three to four researchers attended each session. Participants sat in front of and facing the system. E4 physiological sensors were placed on both participants' non-dominant wrists. Calibration data and a two-minute resting physiological signal were recorded for each participant. Each participant wore a clip-on microphone to record their speech. Nao was positioned in front and to the side of a large TV screen. The Kinect was placed facing the two participants. An administrator operated the experimental workstation placed to the side of the participants. Two webcams were used to record the participants' behaviors and interactions as well as the SAR and VR environment. The details of experimental setup and procedure can be found in Online

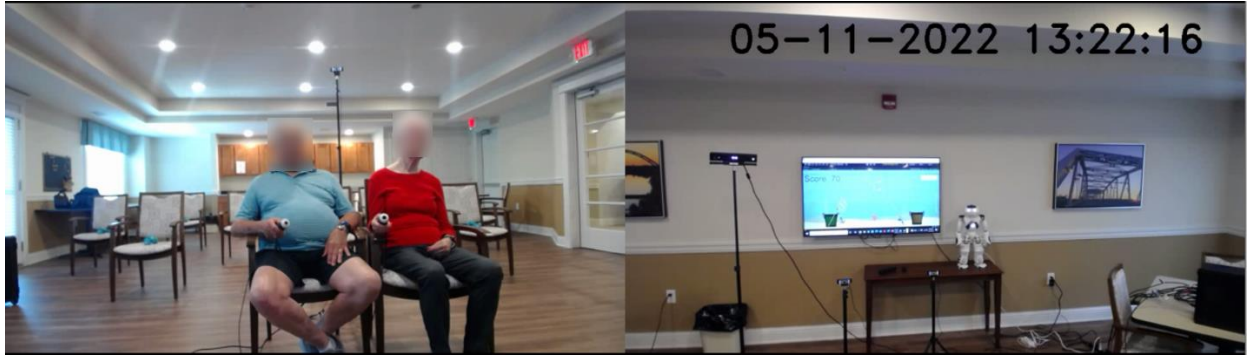

Fig. 4: Left – two participants performing the fishing activity, right – VR screen and Nao in front of the participants; presented with permission from the participants.

Each participant participated in five separate sessions lasting 30 to 45 minutes. Each session encompassed two activities with a 5–10 minute break between activities. At the start of each activity, participants completed an orientation where they learned how to use the wands to map their movements, the objective of the activity, and got the opportunity to practice the basic motions needed for the activity. While most of the sessions were completed with an older adult pair, because of scheduling conflicts some sessions were completed with one older adult and a research partner. After each session, feedback was solicited from the older adults using a structured interview (see Online Resource 4) to measure their degree of comfort and confidence interacting with the system. The participants were also asked several qualitative questions on aspects of activities they enjoyed, the difficulty level, level of engagement, and changes they would like to see. The answers to these questions informed many of the design changes of the activities.

Researchers took field notes during the sessions, and videos were reviewed with the full team to describe the context, technology issues, activity issues, interpretation of issues and develop a plan to address issues. Between each session, engineers modified activities and the SAR system based on the plan.

## 6 Field testing 2

### 6.1 Recruitment of older adults at LTC Site 2

A second LTC site was used for validation testing of the refined prototype activities by a different set of older adults. Recruitment and enrollment for the second sample of participants followed the same screening and consenting procedure discussed in Section 5.1. Eight residents (mean age: 80 ± 4.7, mean AES score: 26.25 ± 2.72, and mean SAGE score: 14.25 ± 4.0) from the second LTC facility were eligible and consented. Table 3 presents participant details. As with Field Testing 1, activities were modified between sessions based on observation and feedback.

**Table 3: Participant details for LTC 2**

| Participant ID | Age | Gender | Education | Technology use | AES-S score | SAGE score | Completed the study |
|---|---|---|---|---|---|---|---|
| A1007 | 82 | M | Bachelor's degree | Everyday | 26 | 18 | Y |
| A1008 | 81 | F | High School | Rarely | 25 | 18 | Y |
| A1009 | 75 | F | Graduate degree | Everyday | 28 | 18 | Y |
| A1010 | 80 | M | Bachelor's degree | 2/3 times a week | 31 | 11 | Y |
| A1011 | 80 | F | Graduate degree | Never Use | 29 | 6 | Y |
| A1012 | 72 | F | Trade School | Everyday | 25 | 12 | Y |
| A1013 | 81 | F | Bachelor's degree | Everyday | 22 | 16 | Y |
| A1014 | 89 | F | Graduate degree | Everyday | 24 | 15 | Y |

### 6.2 System setup by LTC Staff

To determine whether LTC staff could administer the system, 5 staff from Site 2 (mean age 34.2, standard deviation ±15.8) were recruited and consented to the study. Staff participants were asked to set up and run the system based on a written setup guide. All components of the system were labeled and color coded and the labels were referenced in the written guide. Every section had a video guide accompanying the written manual. Only minimal verbal assistance was provided by the researchers when required. After each session, the system set-up and interface were refined based on objective and subjective feedback.

### 7 Results

### 7.1 Usability Measures

At each session, participants at both sites rated their degree of comfort and confidence in a) using the wands, b) interacting with the robot, and c) interacting with the VR environment. Each item was scored on a five-point Likert scale (1 = least positive and 5 = most positive). Between the first and final sessions, the participants at LTC Site 1 showed an improvement in ratings in five of the six categories with the greatest improvement in comfort with the robot. Even though confidence in the VR system decreased by 0.75, the average overall improvement across categories was 0.583. The improvement of user ratings across all categories was statistically significant (p = 0.025) as determined using the Wilcoxon signed rank test. The category-wise mean improvements and standard deviation are shown in Fig. 5.

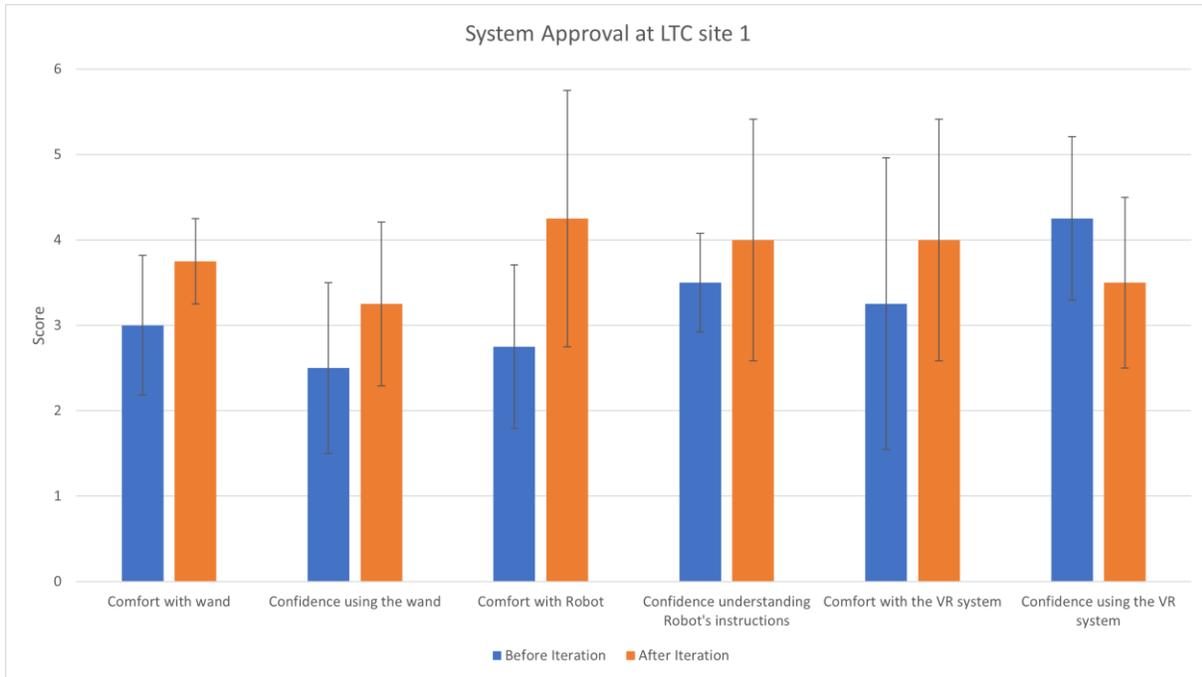

Fig. 5: System approval rating before and after iterative changes at LTC Site 1

The data from the study with participants at LTC Site 2 showed improvement in all six categories with an average of 0.708-point increase per category between the first and final sessions (p = 0.00003). The category-wise mean improvement and standard deviation are shown in Fig. 6.

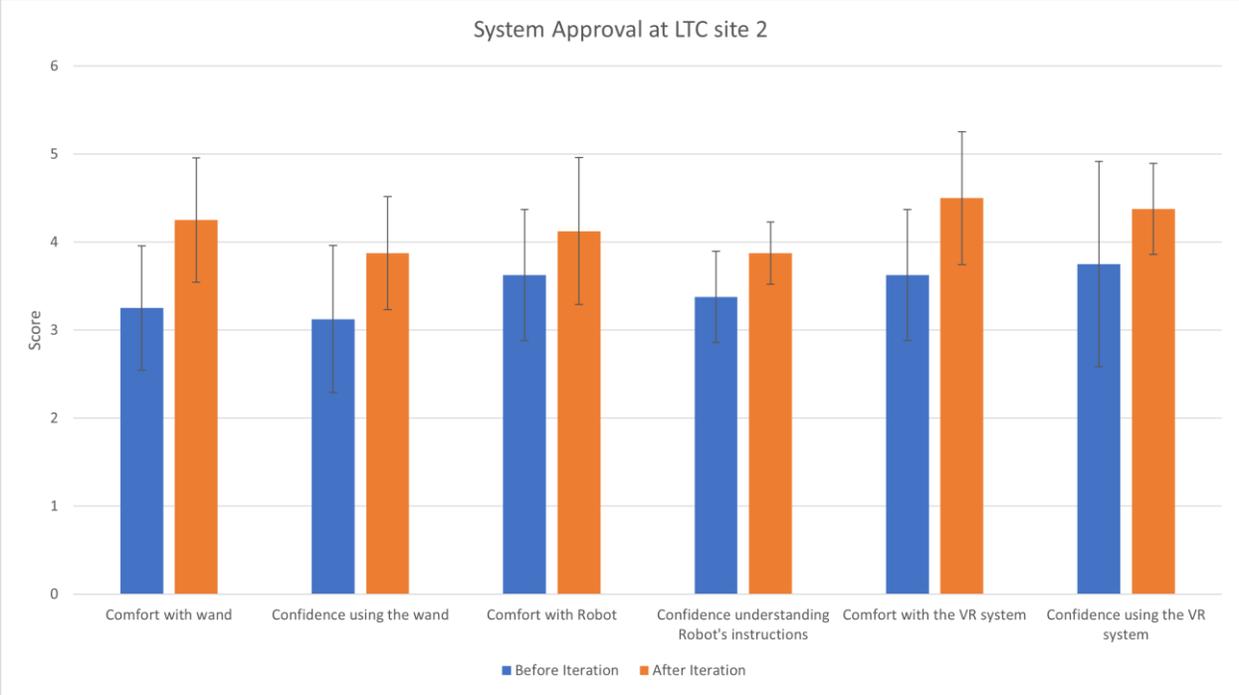

Fig. 6: System approval rating before and after iterative changes at LTC Site 2

Analysis of the recorded video revealed that the average interaction of older adults with their partner increased consistently with iterations from 0.16 interactions per minute during the first set of sessions to 1.14 interactions per minute in the final set of sessions. The same analysis showed the average number of robot interventions decreased from 1.10 per minute to 0.73 per minute. The average number of researcher intervention required also decreased from 0.34 per minute to 0.10 per minute. The results are illustrated graphically in Fig. 7. These numbers suggest an increased engagement and interaction among older adults and increasing effectiveness of the system with each iteration.

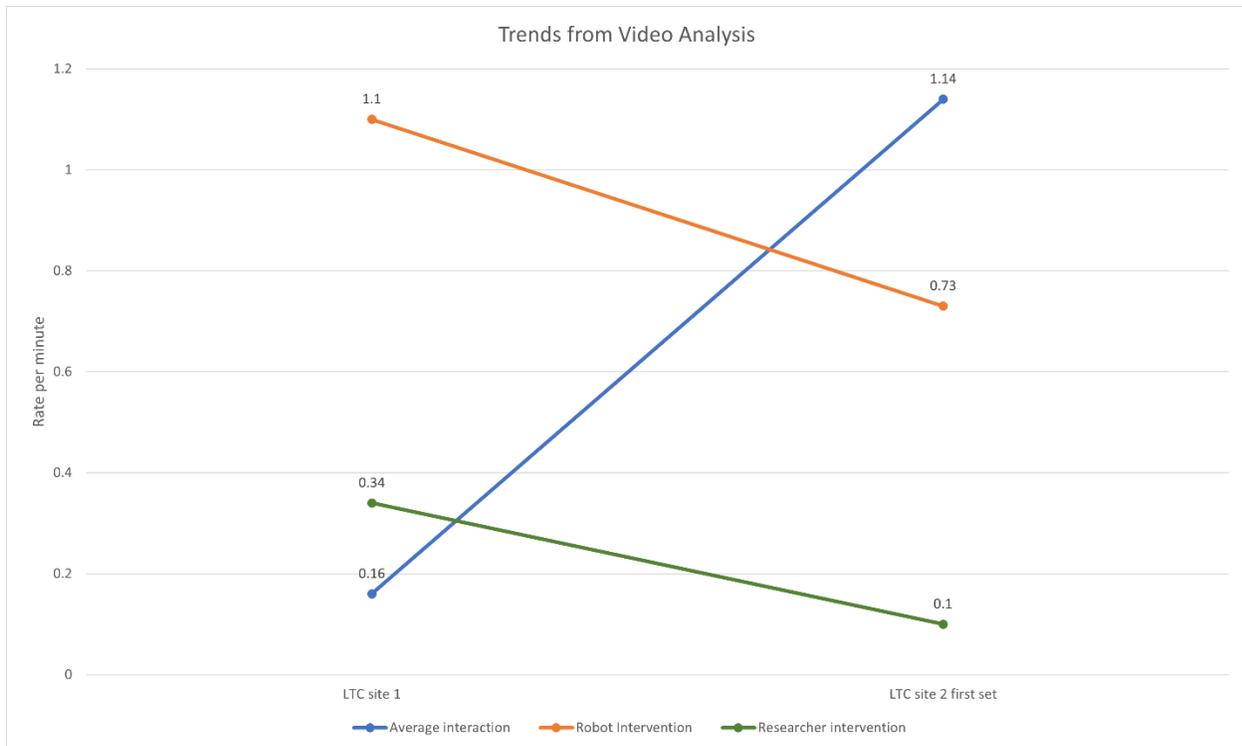

Fig. 7: Results from Video Analysis showing trends in Participant interaction and Robot and Researcher Intervention

## 7.2 Qualitative results

The participants were asked open-ended questions about their experience with the activities and interacting with the robot and they replied that they enjoyed the activities. Some quotes from the participants are mentioned below:

*Research assistant: "Would you like to play these activities again?"*

*Participant 1: "Yes, they were fun."*

*Participant 2: "They were very interesting. I enjoyed doing them."*

*Research assistant: "What did you like about this (fishing) activity?"*

*Participant: "Once I learned how to cast the rod, I really enjoyed it. I think more fish will be fun."*

*Research assistant: "Do you think you will need more instructions from the robot?"*

*Participant: "No, I think the instructions were enough and easy to follow."*

The LTC staff participants were asked if they found the system user-friendly and the instructions easy to follow. They replied that there was an initial learning curve but once they were familiar with the components, it will be easier to set up the next time. A few quotes from the LTC staff are mentioned below:

*Research assistant: "How do you think you will feel if you were to do this (setup the system) a second time?"*

*LTC staff: "Now that I know what everything is, I think I would be a lot better at setting it up."*

*Research assistant: "When you were setting it up, did you find the videos to be helpful?"*

*LTC staff: "Yes. I would read it (instruction manual) through first and try to do it, if I get stuck somewhere then I would go to the video."*

*Research assistant: "In the actual document, was the text easy to follow or pictures will be better?"*

*LTC staff: "I think pictures will help, but I think the text was good. You were very thorough with piece by piece (instructions) about what you needed to do."*

### 7.3  Challenges in the User-Centered Design Process

A major challenge to the UCD process was the ongoing COVID-19 pandemic. The study began in May 2020, at the start of COVID-19. Universities and LTC sites were on lock-down for months. The healthcare system was severely strained, delaying our access to experts from LTC sites. We were unable to conduct initial prototype testing in the VU Engineering Lab as initially planned since older adults were kept in social isolation. The ongoing UCD was interrupted several times as COVID waves continued to occur from 2020 – 2022, restricting access to LTC sites and older adults.

### 7.4  Challenges in Working with LTC Older Adults

Most older adults had mild to significant cognitive deficits impeding their ability to interact with the system or understand the activity. Physical impairments made it difficult to handle the input device. Fatigue occurred easily. Visual impairments impeded their ability to see the VR environment. Hearing impairments impeded their ability to hear Nao's voice. Refinements were made to both activity and SAR-VR system to accommodate these aging changes (See Online Resource 1 and 2).

### 7.5  Challenges in Working within the LTC and LTC Staff

Five LTC workers evaluated the system from the perspective of ease of use and convenience. They set up and ran the system following written instructions prepared by the researchers. After the first two sessions, the staff sometimes had difficulty following the steps in the correct order. To remedy this, the instruction manual was divided into three parts: pre-session, during-session, and post-session. The pre-session consisted of initialization and calibration of components;

during-session procedures involved running the system, activity selection, and changing the levels; and post-session consisted of securely closing all applications and sensors, saving the data, and turning off the system. This division helped avoid confusion and led to a smoother setup. All the staff participants commented that the first-time setup had a learning curve, but they would feel more confident if they were to set up the system again. Overall, they found the integration of the different components of the system to be satisfactory and user-friendly.

Logistics of the UCD proved challenging. Scheduling the sessions was complicated by competing priorities within the LTC, staff illnesses and shortages, and competing use of the activity room. The physical setup took additional time due to internet issues.

### 7.6 Physiology and motion data for future research

Multimodal physiology data including Blood Volume Pulse (BVP), Electro Dermal Activity (EDA), Skin temperature, and accelerometer data were recorded using the E4 sensor, and joint position and orientation data were collected from the Kinect sensor during the sessions. These data along with the performance data recorded by the time-stamped logs will help in further analysis and improvement of the system. Physiological data can provide insights into the affective state (e.g., stressed/ not stressed) of the participants and joint position and head orientation data from Kinect can be analyzed to study participant's range of motion, level of interaction with activity and non-verbal communication with their partner.

## 8 Discussion

In this user centered design study, we found that older adults residing in long term care facilities who had varying levels of cognitive and physical impairments were quite capable of providing important feedback on creating a library of SAR-VR multimodal activities for future studies. Although others [53, 54] have highlighted the importance of user-centered design for older adults, few studies have actually incorporated older adults in the design of technology [36]. Second, it is important to involve LTC staff in UCD such that they feel comfortable in operating new systems and thus ensure system scalability in LTC settings. Thus, we focused on the two key stakeholders: LTC older adults and LTC staff in our UCD of the SAR-VR system and activities. Our findings demonstrated the effectiveness of an iterative user-centered design approach for designing SAR-VR activities and systems for older adults in LTC settings.

The study provided valuable lessons about conducting UCD of SAR-VR systems in LTC settings. First, the timeline for development and testing was prolonged due to multiple waves of COVID-19, virtually shutting down access to the older adults. Second, as is often the case in the design of technology, our initial decisions were based on our expertise and experience, the literature, and experts in the LTC field. Some of these decisions were discovered to be incorrect during field testing. For example, we designed the graphics to be seen easily by the older adults (contrasting colors, font size) and lowered the pitch of Nao's voice to accommodate age-related vision and hearing changes. Nevertheless, we found our graphics required further modifications and a series of voices had to be tested before older adults agreed to the pitch, tone, and speed [55]. We depended on accessing the LTC based on staffing issues and competing demands, further prolonging the process.

Despite these challenges, we found the UCD process successful in adapting the SAR-VR system and multimodal activities for older adults with cognitive and physical impairments. With the increasing population of older adults with ADRD and insufficient skilled care staff, technological interventions are a promising strategy in the delivery of care processes. We developed a SAR-VR system to deliver multidomain activities for older adults residing in LTCs using a user-centered design approach. We designed the activities to provide physical, cognitive, and social stimulus and promote HHI. Such multidomain stimuli along with HHI has been shown to be effective in reducing apathy [11, 13]. Each activity required the participants to cooperate and make joint decisions, thus encouraging interaction and conversations. While other work has attempted to reduce apathy using technology [56, 57], research on using technology to foster and mediate natural HHI among older adults with minimal caregiver intervention is scarce.

Out of the total fourteen participants recruited for the study, twelve stayed for the full duration of the study. Participants' approval ratings across five of the six categories at LTC site 1 increased, while ratings across all six categories increased significantly after the iterative design process at LTC site 2. A closer inspection of the data from LTC site 1 revealed that the rating for confidence in the VR system was already above 4 on a 5-point scale in the first set of sessions. The subsequent decrease might have been a result of the novelty factor wearing off. However, iterations done during the testing at LTC site 2 led to an increase in the rating to 4.4 at the end of the final set of sessions. Overall, the participants enjoyed doing the activities.

Since the system is designed to be deployed at LTCs, we tested the setup procedure of the system with LTC staff to ensure they find it user friendly. All of the enrolled staff could set up and run the system with the provided instructions without any special training, thus proving the practicality and deployable nature of the system in an LTC. When asked if they would be more comfortable if they were to set up the system again, most replied in the affirmative and mentioned that the most challenging part was identifying components that they were otherwise unfamiliar with.

The iterative design process allowed us to make numerous modifications to the various components of the system, including the setup guide, until the system was suitable and enjoyable to both of the target populations, the older adults and the LTC staff.

While UCD has been used to develop e-health applications [58, 59], exercise games [60, 61, 62], and websites [63, 64] for older adults, to the best of our knowledge, this is the first study to take an iterative user-centered design approach to developing a SAR-VR system and activities for older adults. The resultant modular architecture and library of activities sets the foundation for investigating their effectiveness in mitigating apathy among older adults with dementia. The architecture can easily be adapted to other robots, activities, and interaction devices. The intensity of the physical, cognitive, and social components of the activities can be individually tuned, if desired, to study the effect of each component.

Physiological, motion, and performance data collected from the study will provide insights for further analysis and refinement. Physiological data can provide insights into the level of stress and cognitive load of the participants during performance of the activities, which can later be used to design an affective state based closed loop system to automatically regulate the difficulty of the activities. The joint position data from the Kinect sensor can later help automate the detection of non-verbal communication by tracking their head pose and gestures.

## 9      Conclusion

Building upon previous work on SARs and VR in [65], this current work details the use of user-centered iterative design techniques to develop an SAR-VR system and a library of activities that provide physical, cognitive, and social stimuli and promote HHI through cooperation between the participants during the activities. This included four activities: three humanoid robot-based activities, where the humanoid robot Nao acted as the coach to provide hints, corrective feedback, and mediate the interaction between older adults, and one animal robot-based activity, where an avatar provided the feedback and the puppy robot Aibo performed tricks as rewards to incentivize the older adults to complete the activity.

A finite state machine was developed to provide appropriate real-time feedback in response to participants' performance in the activities. The system also collected video, physiology, joint position, and performance data that will be used for future investigations into how stress and engagement patterns in older adults varied with activity difficulty level.

To fulfill the usability requirements, we took a user-centered design approach and gathered feedback from our advisory panel as well as our end-users: the older adults residing in LTCs and LTC staff. We incorporated various tunable difficulty levels in the activities to accommodate different physical and cognitive abilities. The library had a variety of activities using both the humanoid and animal robot to cater to a wide range of participants and prevent monotony. The system had a simple user-interface and easy to follow setup guide for the LTC staff. All of the LTC staff who tested the system were able to complete the setup and run the system without prior training.

Participants' ratings and comments from the LTC staff indicate the usability and acceptance of the system by our target population. This system will be used for a longitudinal study to assess the effect of HCI, HRI and HHI on apathy in older adults residing in LTCs over a longer duration.

The current work, although promising, is not without limitations. It is limited by its small sample size and short duration of the study. A long-term study is required to gauge user perception accurately and mitigate the novelty factor. Also, the user perception of the system depends on several factors like cognitive abilities, previous exposure and familiarity with similar technology, and personal preferences towards particular activities. A larger sample size is required to generalize the preferences as data from smaller samples are easily skewed by outlier individuals.

**ESM1**

**Supplementary Information for
User-Centered Design of Socially Assistive Robotic Combined with Non-Immersive Virtual Reality-based Dyadic Activities for Older Adults Residing in Long Term Care Facilities**

**Details of the SAR-VR System Architecture**

1  Wand design

The wand is a custom designed HCI device that is used to manipulate the virtual environment. From our literature survey on HCI devices for older adults, we found that a controller using both button and motion-based control was the most suitable. Most such controllers available on the market are designed for gaming and not suitable for older adults in terms of weight, grasp, and ease of use. We consulted with an occupational therapist specialized in geriatrics during the design of the wand. The wands feature an ergonomic grip design that can accommodate a variety of palm sizes. The wand dimensions were determined based on guidelines from the Canadian Center for Occupational Health and Safety. The wand is able to control the position of a cursor on the screen, has buttons and a dial for input, that are suitable for older adults. It uses an inertial measurement unit (IMU) to determine its orientation. It also has a vibration motor for haptic feedback. Four versions of the wands were developed over the course of this study; the latest version is shown in Fig. 1. It features an ESP32 microcontroller core from Espressif, (*https://www.espressif.com*), ICM20948 IMU from InvenSense (*https://invensense.tdk.com*), wireless communication, and contact charging.

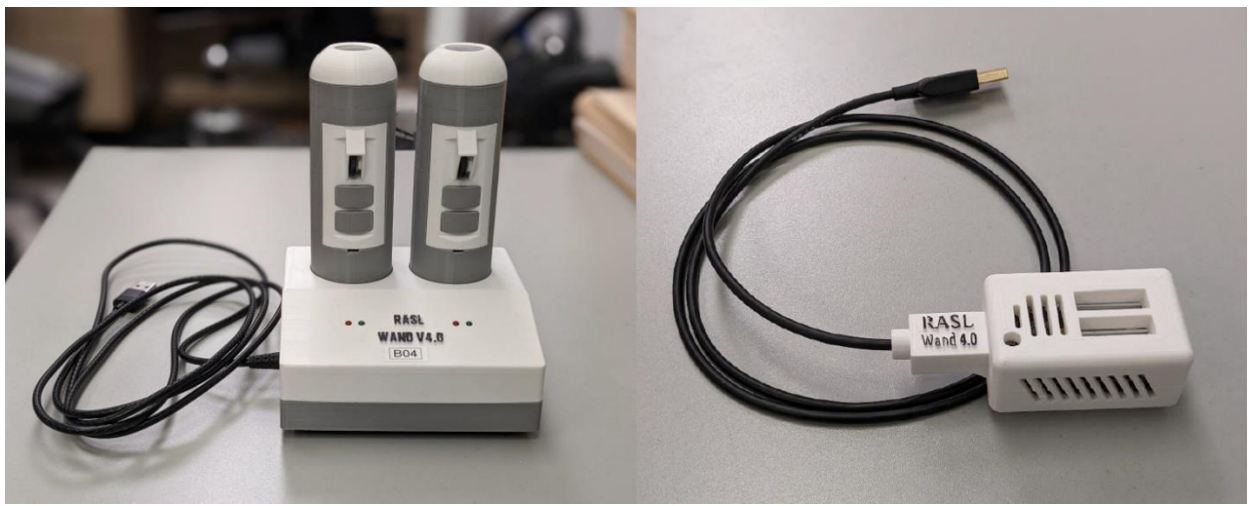

Fig. 1: Left – Wands with charger, right – wireless receiver

2  Finite State machine

Different methods are available for modelling control systems. A Finite State Machine (FSM) was selected as it provides a simple and concise way to structure a control system where the behavior of the system is governed by a set of predefined conditions. An FSM is best suited for modelling systems that exhibit discrete sequential behavior. It has a finite number of states, conditions to transition between the states, and actions associated with each transition. In an FSM model, the system can only be at one state at a time.

When the system first turns on, the state machine is in the "Start State" where preliminary checks are conducted to ensure all components are connected to the system. Once verified, the system enters the "Stable State". This state conducts the regular flow of the activity. Each activity has a set of defined states and corresponding state transition conditions. Fig. 2 shows the block diagram of a state machine for the Music Activity.

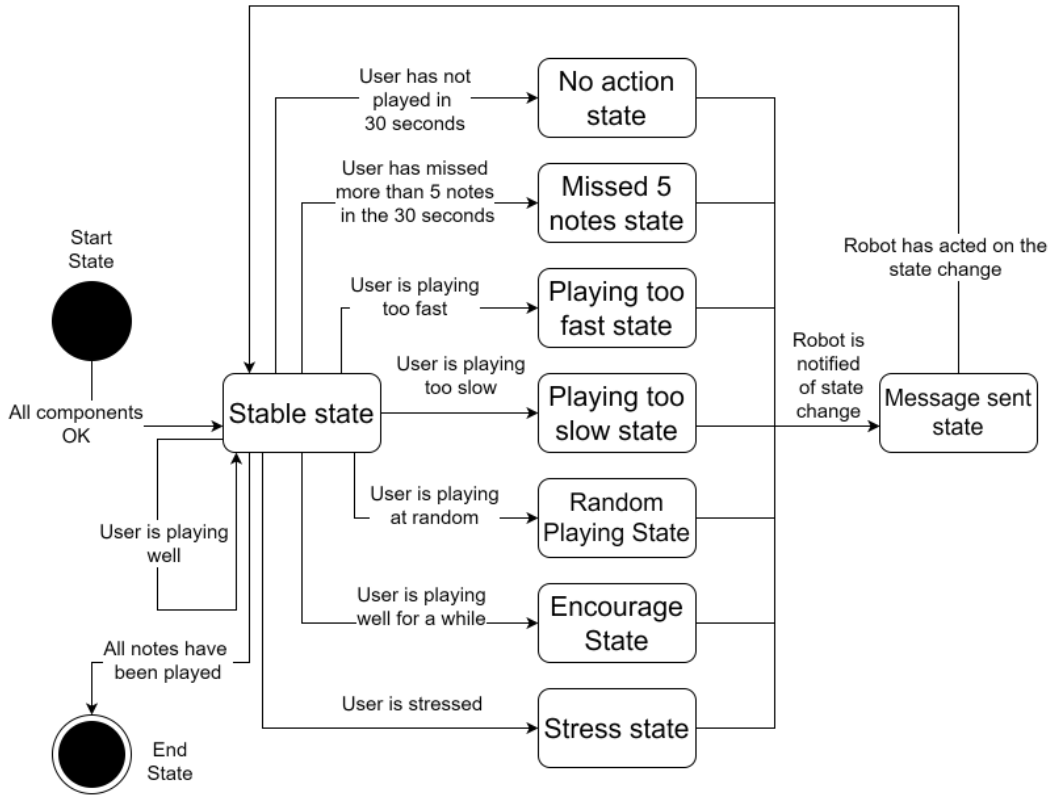

Fig. 2: Block diagram of state machine for Music Activity

## 3  Robot communication layer

The system was developed to be flexible with various makes of robots. Feedback can be conveyed from any virtual avatar or robot and will only require changes to one intermediate layer. This intermediate layer is used to translate human-understandable, English messages such as "LeftPlayingFast" to robot-specific messages. This design allows for easy exchange of the type of robot used with minimal change in software. Only the robot side and the communication layer will need to be modified and the activity itself will not need any modification. For example, if Nao needs to be replaced by some other humanoid robot, the state machine output will remain the same.  The communication protocol supported by the new robot will be added to the communication layer and the functional behaviors will have to be installed in the new robot. Fig. 3 shows the block diagram of the communication layer.

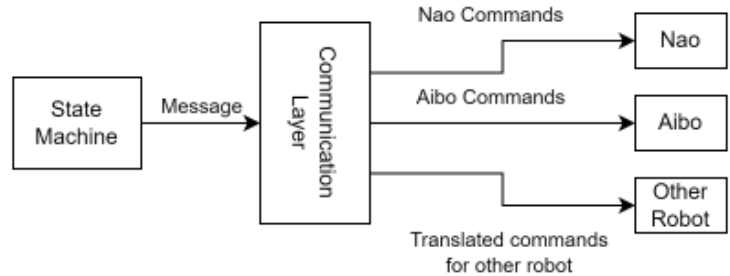

Fig. 3: Block diagram of communication layer

## 4  Feedback animations and speech

The humanoid robot Nao is used to provide instructions and feedback to the participants during the activities. During the initial design of activities and meetings with older adults, multiple voices were generated for Nao using play.ht (https://play.ht), an online AI speech synthesizer. The pitch, speed, and pace of speech were adjusted until the older adults provided a consensus on the preferred speech domains. The final parameters chosen are mentioned below.

Voice: Guy, Male, English

Voice Style: mix of Regular, Cheerful, Excited, Friendly

Frequency: 48 KHz

Speed: 80%

Nao's animations and speech fall into three major categories: instruction, corrective feedback, or celebrations. Instructions are typically given during the tutorial and throughout the activity if a new component is introduced. The corrective feedback is prompted when a participant is either not completing part of the activity or is struggling to complete the required action. Corrective feedback is designed to always be positive and encouraging while providing reminders or hints to the participants. Each activity has activity-specific feedback and celebrations. There are positive encouragements that are interspersed throughout all activities. Each speech component has corresponding movements programed for Nao using Choregraphe (www.softbankrobotics.com). The movements are designed to match what is said and to mimic typical human body language. If Nao is directing the participants' attention to the screen, it will use its arms to gesture towards the location of the element it is referencing.

Aibo has several built-in tricks and functions from Sony (us.aibo.com), such as sit, dance, and rollover. Using the web API, the state machine can call specific actions to have Aibo perform tricks during the dog related activities.

## 5    Data collection

### 5.1    Video recording

Video data were collected to examine participant behavior and response to robot interactions (instruction, feedback, and celebration). Two cameras were used, one to record the participants and the other to record the robot. A custom video recording tool was created using Python3 and OpenCV (https://opncv.org) to simultaneously record two monitors and two cameras with synchronized audio and time stamps. This composite video can be used to label participant behavior for future machine learning applications and for diagnosing system issues. Recordings were coded using the Noldus Observer XT® system (Noldus Information Technology, Netherlands) for analyzing behaviors utilizing a coding scheme to identify dyadic, human-computer, and human-researcher interactions.

### 5.2    Performance data

The system generated time-stamped text logs recording the performance of the participants on the activities, the response of the system and robot feedback. These logs can be analyzed to gain further insight into the performance of the participants and check the effectiveness of the system.

### 5.3    Joint position and posture data

The Kinect v2 sensor from Microsoft is used to capture the joint position of each user through a Python program. For each user, we are able to capture 32 different joint orientations and positions with a sampling rate of 30: Pelvis, Spine navel, spine chest, neck, clavicle left, shoulder left, elbow left, wrist left, hand left, hand tip left, thumb left, clavicle right, shoulder right, elbow right, wrist right, hand right, hand tip right, thumb right, hip left, knee left, ankle left, foot left, hip right, knee right, ankle right, foot right, head, nose, eye left, ear left, eye right, and ear right. The orientation and position information of each joint is saved to a JSON file where the keys are the timestamp at which the sample was collected.

At each sample, the program scans the number of people in the frame and then collects joint information about each. By default, the Kinect sensor can capture information up to 4.5 meters away; we limited the range to 2.5 meters as the users will be sitting close to the sensor. This allowed us to ignore people walking in the background. Additionally, if there are more than two people detected within the range of interest, the sensor stops collecting information. With only two people in the frame, the program can differentiate between the two users through their relative positions.

### 5.4    Physiology data

The E4 sensor from Empatica (https://www.empatica.com) was used to collect physiological data. The E4 is a research grade wearable wristband that collects data from a photoplethysmography (PPG) sensor, an electrodermal activity (EDA), an infrared thermopile, and a 3-axis accelerometer. From the PPG, the E4 outputs heart rate data, blood volume pulse, and heart rate variability. The raw EDA signals can be decomposed into relevant features such as skin conductance response and skin conductance level. Body temperature is available from the thermopile and accelerometer data in 3 directions are given from the accelerometer. All data are saved in individual csv files.

In the context of this work, the goal was to ensure that physiological data could be successfully collected during the activities with minimal losses of data. Future work will explore the use of physiological data and machine learning to create predictive models for stress and cognitive load.

## 6      Menu system

In order to make the SAR-VR system acceptable to the LTC staff, it must be easy to operate. Hence, we have designed a graphical user interface for the LTC staff. When the system is launched, the LTC staff member is presented with a simple multi-page menu system. On the first page, the names of the two participants are entered to allow the robot or avatar to provide appropriate feedback directly to each participant. The staff member chooses between the humanoid or animal SAR-based activity. If the humanoid SAR activity is chosen, the IP address of the robot will be requested to ensure that the system is communicating with the correct robot. Finally, a specific SAR humanoid activity, such as Music, is chosen with a designated difficulty level (Level 1, Level 2, or Level 3). At that point the activity begins. If the animal robot is chosen, no IP address is requested as the communication to the Aibo robot is handled through an API key.

The menu system was developed to decrease the number of manual steps, limit human burden, and reduce errors. For example, the wands communicate with the computer using communication ports similar to other peripheral devices. Instead of requesting the LTC staff member to search through multiple communication ports connected to the computer and manually inputting into the menu system, the menu system automatically selects the correct port by comparing the output of the ports to the expected output of the wand. Fig. 4 shows example pages from the menu system.

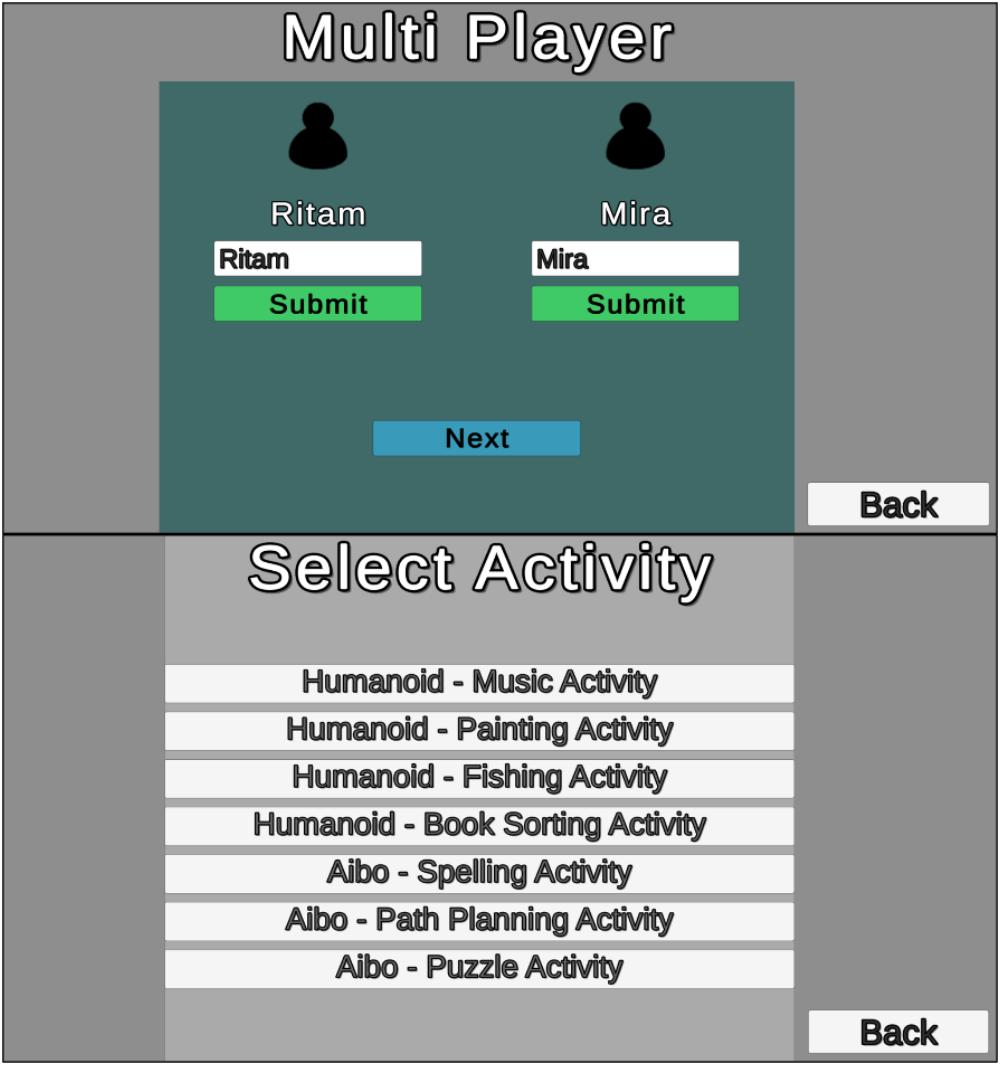

Fig. 5: Top – Player selection screen, bottom – Activity selection screen

# ESM2

# Supplementary Information for
# User-Centered Design of Socially Assistive Robotic Combined with Non-Immersive Virtual Reality-based Dyadic Activities for Older Adults Residing in Long Term Care Facilities

**Activity design**

We explored potential virtual activities with the LTC activity directors and older adults residing in LTCs, guided by the framework of multidomain activities that incorporate physical, cognitive, and social stimuli. Based on the conversations that took place during the brainstorming sessions, inspiration for activities was taken from real-life activities that the older adults perform at the LTCs and hobbies they used to enjoy in their youth. Among the potential activities discussed, we began with four initial activities:

1. A *music activity* was designed that requires the participants to play virtual drums along with music from their youth.
2. A *fishing activity* was designed where they can catch virtual fish in a virtual lake environment.
3. A *painting activity* was designed that allowed the participants to paint pictures by numbers.
4. A *spelling activity* was designed that allowed the participants to train Aibo by spelling out dog commands by selecting virtual letters.

# 1      Music activity

The music activity utilizes gross motor movement by asking users to perform a drumming motion with the wand to play virtual drums in sync with displayed music notes. It involves cognitive skills of perception and attentional control. It encourages HHI by performing with another person. A menu of songs was provided based on suggestions from the LTC staff and the older adults and included various genres, such as swing, hymns, country, and rock-n-roll.

A software-based audio spectrum analyzer isolates the drumbeats from the music, and musical note symbols are generated in synchronization with these beats. These notes travel down the screen along two vertical bars, the left bar for the left participant and the right bar for the right participant, as can be seen in Figure 1. The participants earn points by hitting the drum when the note is crossing a green zone in the bar. Both participants have to perform well to maximize points, which incentivizes participants to encourage one another. In addition to the green zone, higher levels of this activity have a yellow zone above and a red zone below the green zone that indicates the note being played too early or too late, respectively. These extra zones introduced an added layer of cognitive difficulty. NAO guides the participants through the tutorial and provides encouragement and corrective feedback during the activity to increase engagement.

# 2      Fishing activity

The fishing activity utilizes both gross and fine motor skills, as well as problem solving, attention, and working memory. Furthermore, in order to catch the fish and successfully transfer the fish to the bucket, the two participants must coordinate and cooperate to accomplish their shared goal. In the final version of the fishing activity, the participant on the right side controls the fishing pole. The participant makes a casting motion with the wand, which casts a fishing line onto the screen depicting a virtual world of water with a school of fish. At the end of the line is a cursor controlled by the movement of the wand. Using this cursor, the participant can catch a fish. Once captured, the participant on the left must use their wand to control a net. The two participants must coordinate to transfer the fish from the rod to the net. After the fish is successfully transferred, the left player is able to deposit the fish into a virtual bucket to score points. During the activity, NAO provides reminders for next actions, as needed. The feedback is prompted by the state machine if the older adults take too long to perform their steps. Nao also encourages them to seek help from their partner, if required and provides positive reinforcement throughout the activity.

## 3   Painting activity

The painting activity focuses on fine motor skills to precisely guide a paintbrush to the correct segments in an image, while also employing recognition and attention. To complete the full painting, participants must collaborate.

The activity is structured as a paint-by-numbers format where each participant has a set of assigned numbers and is only able to paint the segments corresponding to their numbers. Each participant has a color palette that contains the colors corresponding to the segments they are painting. Using a virtual paintbrush, each participant can choose a color and then guide the paintbrush to the appropriate segment to fill in the color, as shown in Figure 1. If a participant faces difficulty performing the activity, NAO provides helpful reminders about the next steps, as well as encouragement. In more complex levels, each participant has an increased number of segments to paint.

## 4   Spelling activity

The spelling activity focuses on fine motor skills for cursor control and cognitive skills to recall and spell a word presented to the participants. This activity uses the dog robot Aibo to entertain the participants upon activity completion and serve as a reward to keep them engaged.

In the virtual environment, a word that corresponds to a dog command (e.g., sit, shake, dance, etc.) is given to the participants. Using the wand, the participants are instructed to choose letters to spell out the command that was prompted. Half of the letters are randomly colored red; the others are blue. Each participant can only choose one color of letters, creating collaboration within the activity. After each word is spelled out, Aibo performs the corresponding trick as a reward. In this activity, a screen-based avatar, shown in the bottom right of Figure 1, gives instructions and feedback instead of Nao to separate the effect of the two types of SAR.

Tables 1 and 2 show the changes made to the activities and the rationale behind the changes.

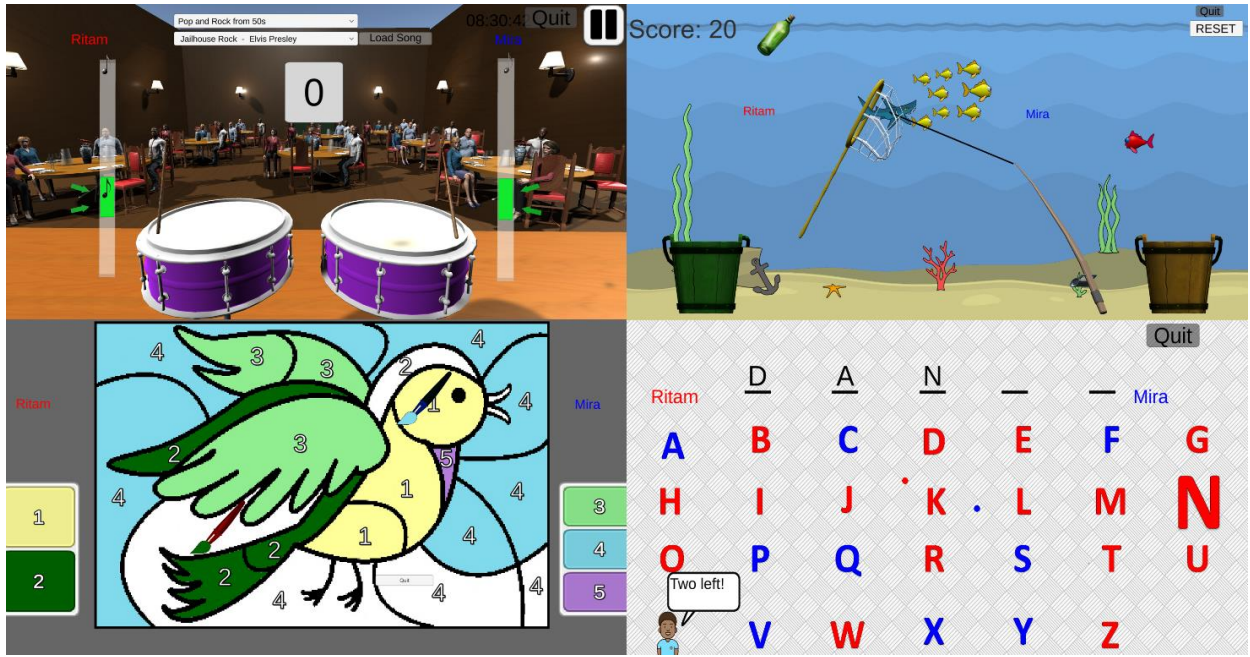

Fig. 1: Top left – Music Activity, top right – Fishing Activity, bottom left – Painting Activity, bottom right – Spelling Activity

**Table 1. Iterative Design Changes to Activities Based on Initial Field Testing with Older Adults**

| Component | Changes | Rationale for changes |
|---|---|---|
| Music Activity | Note generation changed to 'alternate' instead of 'random'. | Sometimes only one participant got consecutive notes while the other had to wait for extended periods. |
| | Adjusted time between NAO's feedback. | Sometimes the consecutive feedbacks were too frequent and overwhelming. |
| Fishing Activity | The initial activity of two fishing rods was replaced by one fishing rod and one net. | The activity now required the participant with the rod to catch a fish and pass it to the participant with the net, who deposited it into the bucket. This increased collaboration and communication. |
| Painting Activity | The cursor changed from a 'dot' to a 'paint brush' with the tip changing color to reflect the color that was selected. | Participants had difficulty seeing the color of the dots and to know which color was selected. |
| Spelling Activity | A reminder was added to cue participants when it was their turn to pick a letter. | Some of the participants did not realize it was their turn to pick the letters and would keep waiting. |
| | | Sometimes, the participants forgot what the target word was. |

| Component | Changes | Rationale for changes |
|---|---|---|
| | A hint was added to remind both participants of the word they are supposed to spell. | Some participants had difficulty locating the correct letters and reducing the number of excess letters helped reduce visual stimuli overload. |
| | A slider was added that can control the number of excess letters visible on the screen. | |

**Table 2. Summary of Activity Changes and Rationale During Testing at LTC Site 2**

| Component | Changes | Rationale for changes |
|---|---|---|
| Music Activity | Alternate notes replaced by a probability approach | The alternate notes for the left and right participants made the activity too predictable and easy. |
| Fishing Activity | Only one bucket active at a time in the highest difficulty level | This required the participants to comprehend instructions more clearly and added to the cognitive stimulus. |
| Painting Activity | Color palette changed from a two-column layout to a single column layout | Sometimes the participants unintentionally switched colors while crossing over the inner palette. |
| Avatar for Dog tasks | Instructions for dog tasks given by NAO were replaced by an avatar | Keeping the two types of SAR separate allows us to compare engagement based on a) activity chosen and b) type of SAR. |

# ESM3

# Supplementary Information for
# User-Centered Design of Socially Assistive Robotic Combined with Non-Immersive Virtual Reality-based Dyadic Activities for Older Adults Residing in Long Term Care Facilities

# Setup Guide

This document will contain a full set up guide for all the sensors, hardware, and software we will be using.

## Contents

- Room Setup Diagram

*This is a suggested setup diagram and will need to be adapted depending on each site space. For example, the Researcher Station may need to be setup on the opposite side of the diagram so that the wires and setup are not blocking the door. Use your best judgment in setting up and use this diagram as reference for where to position things in relation to each other.*

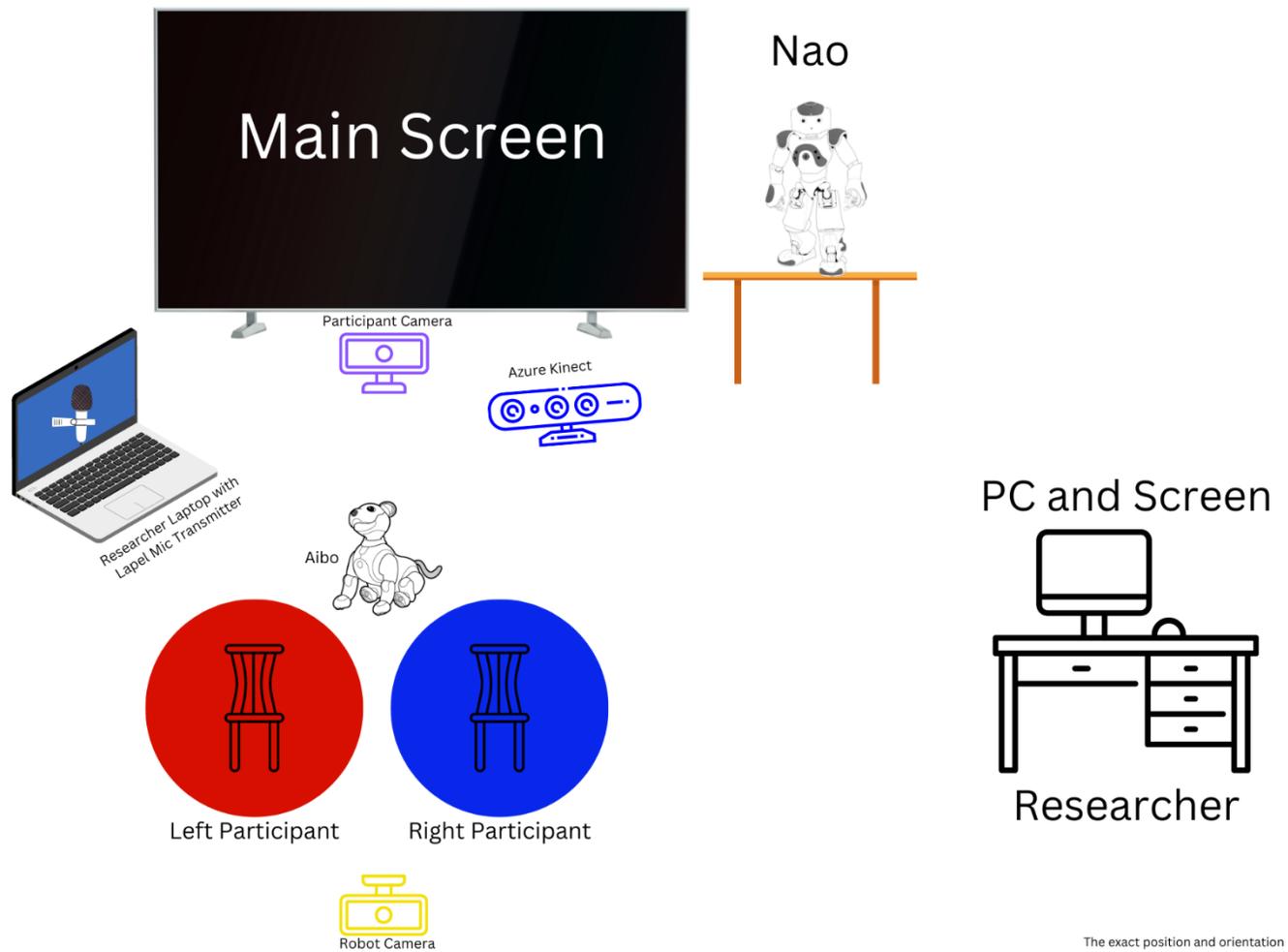

- Pre-Session Setup
- Computer Setup

**Tools needed:**

☐ Computer
☐ Computer power cord
☐ Mouse and Keyboard
☐ Wands
☐ Monitor
☐ Monitor power cord
☐ 3ft Mini Display Port (MDP) to HDMI Cable
☐ 10ft Mini Display Port (MDP) to HDMI Cable
☐ 25ft Power Extension Cable
☐ Power Strip/Surge Protector
☐
☐
☐
☐ **Steps to set up the computer:**
  ☐ Find the Computer Power Cord (labeled with a red label) and connect it to the computer
  ☐ Plug the power cable into the surge protector and plug the surge protector into a power socket (use the power extension if needed)
  ☐ Connect the mouse and keyboard to the computer using the USB cables
  ☐ Find the Monitor Power Cord (labeled with a red label) and connect it to the monitor
  ☐ Plug the screen power cord into the surge protector/power strip
  ☐ Find the 3ft and 10ft MDP to HDMI cables (labeled with orange labels)
  ☐ Plug the MDP side of the 3ft cable into the computer (labeled with computer side)
  ☐ Plug the HDMI side of the 3ft cable into the screen (labeled with Screen Side)
  ☐ Plug the MDP side of the 10ft cable into the computer (labeled with computer side)
  ☐ Plug the HDMI side of the 10ft cable into the screen being used for the task (projector, tv, etc)
  ☐ Power on the computer and TV and log in to the computer
  ☐ this site may need a hotspot depending on if the site's wifi has firewalls that block the robot's codes. If this is needed, please plug in and power on the hotspot now to allow time for the computer and robot to connect to the hotspot.

During this time, it is a good idea to plug in the wands, E4s and the dog, as they need time to charge prior to the session. <mark>Please allot 1 hour in scheduling for set up before the participant comes.</mark>

- Alternate Laptop and Lapel Mic Setup

**Tools needed:**
  ☐ Work Laptop and Laptop Charging Cable

☐ Lapel Mics, Receiver, and Transmitters
☐ AUX to USBC adapter for Dell Latitude 5520 (if needed)

**Steps to set up the Laptop and Lapel Mic:**

☐ Plug in the laptop to the closest outlet on a surface that is close to the participants, Ideally between the participants and the speakers of the TV and NAO

☐ Plug the smartphone adapter (black side) into the left side of the receiver (the one that looks like headphones)

☐ Plug the gray side of the smartphone adapter into the USB C adapter

☐ Plug the USB-c adapter into your computer

☐ Open Sound Settings in Windows and Set the <u>output</u> to your <u>Realtek audio</u>

☐ Set the <u>input to the USB input</u>

☐ Put the microphones on the participants

☐ Record video using camera app on computer

☐ After session, upload video to Sharepoint

- Router Setup (If Needed)

**Tools needed:**
- ☐ Router
- ☐ Router power cable
- ☐ Ethernet Cable

**Steps to set up the router:**
- ☐ Connect the power cable to the router and the surge protector
- ☐ Find the Ethernet Cable (Labeled in Orange)
- ☐ Plug the ethernet cable into the blue internet port on the back of the router
- ☐ Plug the other side of the ethernet cable[1] into the facilities ethernet port (typically found on the wall)

- Robot Pre-Session Setup

*(only one robot at a time)*

- Nao:

**Tools needed:**
- ☐ One Nao Robot

**Setting up Nao**
- ☐ Remove Nao from his box and set him on the table in the squatting position shown below

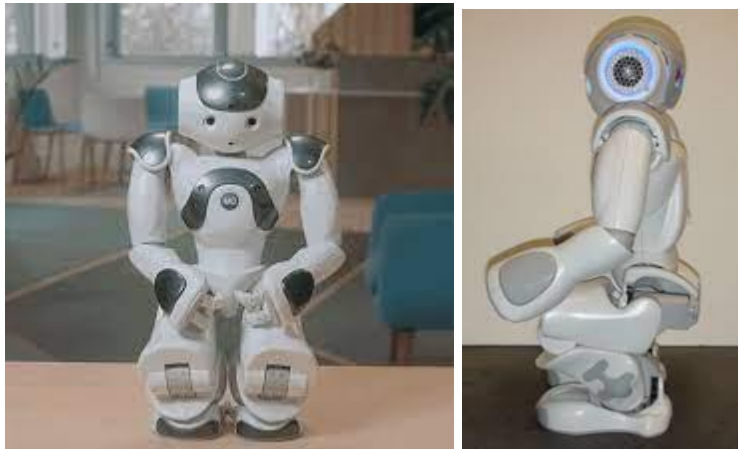

- ☐ Turn on Nao by pressing the button in the center of the robot's chest **(do not hold the button)**
  - o Stay near Nao as he powers up to make sure he is balanced and does not fall
  - o Nao will begin to power on, the lights will turn on and it will stand up. Please wait until he is fully standing to proceed
  - o Plug in the robots power cable if needed (Nao will tell you out loud if his battery needs charged)

---

[1] You may need to talk to the facility in order to activate the ethernet port in the testing room. Please plan to have at least one setup day in each facility prior to beginning testing with residents The system will not work without the use of an active ethernet port.

Page **46** of **55**

## OR

- Aibo:

**Tools needed:**
☐ One Aibo Robot

**Setting up Aibo**
☐ Place Aibo on the floor in the position shown below

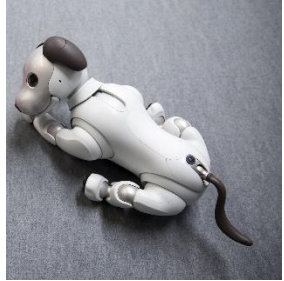

☐ Press and hold the button on his neck until the light turns green
- o Aibo will begin to stretch as he wakes up. He is ready to go when he barks twice in a row

- E4 Setup

**Tools needed:**
☐ 2 E4 Devices with Chargers

**Prepping the E4:**
☐ Find the E4 Sensors (they are light blue boxes with labels on them). One sensor will be labeled red and the other will be labeled blue.
☐ Wipe down the E4s with disinfectant
☐ You can plug them into the computer to charge while waiting for the participants to arrive.

- Video Recorder Setup

**Tools needed:**
☐ 1 webcam labeled as Participant Camera with a Purple Label
☐ 1 webcam labeled as Robot Camera with a Yellow Label
☐ 1 USB Extension labeled with a Purple Label
☐ 1 USB Extension labeled with a Yellow Label

**Setting up the recording system:**
☐ Find the webcams labeled Participant Camera and Robot Camera and mount them both on tripods
☐ Place the camera marked 'Robot Camera' behind the two participants facing the robot. Use the diagram as reference
- If the webcam cord is not long enough, use the USB extension cord (labeled with a yellow label)

☐ Place the camera marked 'Participant Camera' facing the participants next to the screen. Use the diagram as reference

- If the webcam cord is not long enough, use the USB extension cord (labeled with a purple label)Click here to enter text.
- Kinect Setup

**Tools needed:**

☐ Azure Kinect Sensor
☐ Tripod Mount
☐ Kinect Wires (Kinect power cord and USB cable)
☐ USB Extensions (Data extension labeled with a Blue Label, Power extension labeled with a Red Label)
☐ Computer
☐ Kinect Program

**Setting up the Kinect sensor:**

☐ Mount the Kinect on a Tripod
☐ Set up the Kinect sensor on a stand facing the participants, to the left or right of the monitor or projector, 6 - 7 ft away from the participants
☐ Plug the provided USB cable into the back of the Kinect
- If the USB cord is not long enough to reach the PC, use the USB extension cord (labeled with a blue label)

☐ Use the provided power cable to connect the Kinect to power
- Connect the power USB to the power block and connect it to the surge protector
- If the power cable will not reach the surge protector, use the USB Power extension cord (labeled with a red label). Remove the USB from the Kinect power brick, insert the USB extension, and reconnect the power brick.

- Wand Setup

**Tools needed:**

☐ 2 Wands

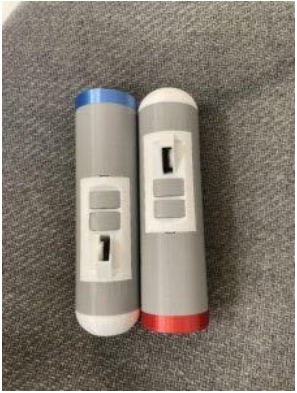

☐ 1 Wand Charger

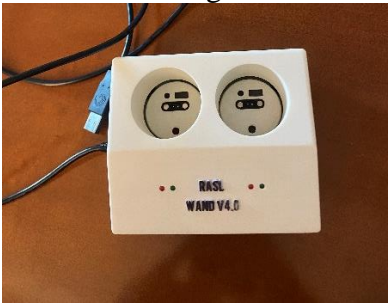

☐ 1 Wand Receiver

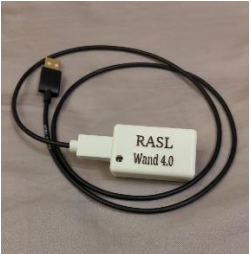

**Prepping the Wands:**

☐ Find the two wands. One wand will have a red bottom and the other will have a blue bottom.

☐ Wipe down the wands with disinfectant

☐ You can plug the charger into the computer and charge the wands while waiting for the participants to arrive.

☐ Plug in wand receiver, this is imperative to wands working, session will not run without this step

- Session Setup (When Participants are Present)
  - E4 Part 1
    - ☐ Unplug the E4s and set the chargers aside for the post session steps
    - ☐ Put the E4 with the red label on the non-dominate wrist of the left participant and turn on (red recording light)
    - ☐ Put the E4 with the blue label on the non-dominate wrist of the right participant and turn on (red recording light)
    - ☐ Let the E4 sit on the wrist for up to five minutes before turning on the device to allow the sensors to warm up to body temp
  - Video Recorder
    - ☐ Plug both the Robot Camera and the Participant Camera into the PC
    - ☐ Double click on the **OBS** icon on the right side of the Desktop
    - ☐ Make sure there are no old videos in the Video folder of the PC
    - ☐ Press Start Recording
  - Kinect
    - ☐ Use the USB cable to connect the Kinect sensor to the computer (USB 3.0 port must be used, those are the blue colored USB ports).
    - ☐ Set up the Kinect sensor on a stand facing the participants, to the left or right of the monitor or projector, 6 - 7 ft away from the participants
    - ☐ On the right side of the computer desktop, double click on the **Kinect** Icon
    - ☐ Press preview button and make sure stick figures of both participants are visible. Adjust the Kinect as needed.
      - Press the preview button to make the preview disappear
    - ☐ Press Start Recording
    - ☐ Press the Windows Key + d to return to the desktop
  - E4 Part 2
    - ☐ Turn on the E4 by pressing and holding the button for 2 seconds
    - ☐ Wait approximately 1 minute for the light on the E4 to turn red indicating that the device has entered recording mode (the light may turn off to save power, just press the big button one time to see the light again)
    - ☐ <u>Collect at least 3 - 5 minutes of baseline before the task begins</u>
  - Wands
    - ☐ Take the Wands off Charge and unplug the Charger if the USB space is needed
    - ☐ Plug in the Wand Receiver into the computer
    - ☐ Turn on the Wands using the switch on the bottom and hand the corresponding wands to each participant
      - The left participant will get the Red Wand and the right participant will get the blue wand

<u>Meaning of Colors on Wand</u>

Red: Wand is charging

<span style="color:red">Red</span> and <span style="color:green">Green</span>: Wand is almost fully charged but is still charging

<span style="color:green">Green:</span> Wand is fully charge

- Socially Assistive Robotics Application
- Steps for the computer:
    ☐ Double click the **Socially Assistive Robotics** icon on the desktop to run the program
    ☐ When you reach the start menu, choose **Multiplayer**
    ☐ Enter the participant's names in order (name of participant sitting on the left facing the screen should go to the left box and the other participant's name on the right box) and hit submit after each name, press **Next** once the names are submitted
    ☐ Choose the activity the participants would like to use
    ☐ Choose the activity level the participants want from dropdown menu
    - o Level 1 is the Tutorial
    - o Levels 2-4 are the main levels with 2 being the easiest and 4 being the most difficult
        - Find more activity specific information on page 13
    ☐ <mark>DO NOT</mark> press Connect or Start. Move onto the robot section below.

- If doing a Nao Based Activity (Music, Painting, Fishing, or Book Sorting)
  ☐ Get the IP address by pressing the button on the chest once (do not hold, just press)
    o Enter Nao's IP address in the. Add periods after each number Nao says (e.g., if Nao says '192','168','1','2', the IP address is: 192.168.1.2). The IP address goes in the box immediately below the level selection dropdown. See picture below.

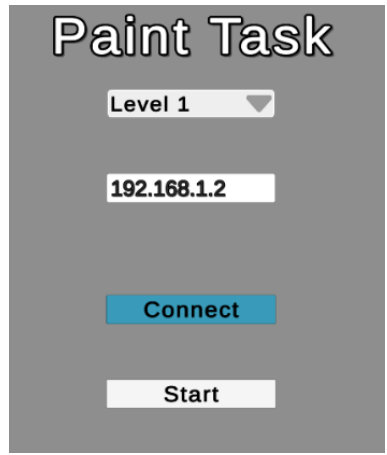

  ☐ Press Connect on the screen. The Robot will do a squat motion and stand up, once he is done, move to the next step
  ☐ Make sure the wands are on and the wand dongle is plugged in
  ☐ Press Start on the screen to being the activity

- If doing a Aibo Based Activity (Spelling, Path Planning, Puzzle)
  ☐ If Aibo is on and ready, place him on the floor in front of the participants.
  ☐ Make sure the wands are on and the wand dongle is plugged in
  ☐ Press Connect on the screen.
  ☐ Press Start on the screen to being the activity

- Activity Specific Information
  **Music Task:**
    o Level 2 is a free play. No notes will appear, the participants can just play the drum as they want
  **Spelling Task:**
    o The Dog task has a slider that controls how many additional letters show up. If you move the slider all the way to the right, only the letters that are need for the word will appear.

- Post-Session
- Before the Participants leave the Room
    ☐ Remove the E4s from the participants and turn them off by pressing and holding the button
    ☐ Thank the participants for their time and ensure that they receive their gift cards
- After the Participants Leave the Room

*Socially Assistive Robotics Application*
☐ If the application is not already closed, hit quit and then close out of the application.

*Video Recorder*
☐ Press Stop Recording
☐ Close OBS

*Kinect*
☐ Press Save
☐ Once the data is saved successfully, close the application window
***Once there is a final "recordings saved" textbox you can unplug all cameras to plug in E4s***

- Data Uploading
    ☐ Connect the E4s to the computer using the dock and cord provided in the box
    ☐ Open the **E4 Manager** on the desktop and press sync sessions for both devices
    ☐ After syncing, press view sessions
    ☐ Press the little box with arrow on right side of the screen to open E4 connect and access the data, click on **Sessions**.
    ☐ Download the two session zip files (match the timestamps and device ids)

    ☐ Click the **DataManager** icon on the desktop
    ☐ Once you enter the full facility id (has to match on Sharepoint data storage folder name (i.e. 003)) and the IDs of both participants, a file will be created on the desktop with the facility name, participants IDs, and the current date
    ☐ Select what you want to happen with the data from the options:
        ☐ Upload to SharePoint (will require username and password)
        ☐ Upload to secure drive
        ☐ If nothing is selected, the zip file will be created on the Desktop, no upload
    ☐ Press the continue button
    ☐ ALSO after session, upload extra videos video to Sharepoint

    ☐ Do not exit out of either page until the file has been fully uploaded
    ☐ At this time on the laptop search" camera roll" to open your camera roll folder and delete any videos from the session as well as deleting them from your laptop. These videos should not leave the site once they are recorded and uploaded

# ESM4

# Supplementary Information for
# User-Centered Design of Socially Assistive Robotic Combined with Non-Immersive Virtual Reality-based Dyadic Activities for Older Adults Residing in Long Term Care Facilities

# Questions to be asked after **each task:**

**Open-Ended Questions:**

1) What did you enjoy about this task?
2) Is there anything you think we should add or remove from the task?
3) Would you like to do this task again in the future?
4) Did you understand the task? Did you need more instructions?
5) How could we make the task easier?
6) How could we make the task harder?

# Questions to be asked at the **end of each visit:**

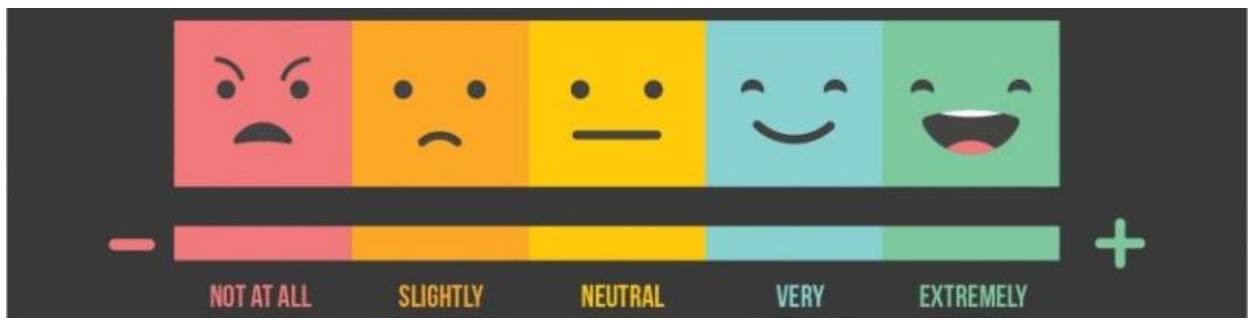

**Using the chart above,**

1) How would you rate your *comfort* level with the wand?
    a. Example answer: I was not at all/extremely comfortable with the wand.
2) How would you rate your *confidence* level with the wand?
    a. Example answer: I was neutral confident with the wand.
3) How would you rate your *comfort* level interacting with the robot?
4) How would you rate your *confidence* level interacting with the robot?
5) How would you rate your *comfort* level interacting with the screen?
6) How would you rate your *confidence* level interacting with the screen?

**Open ended Questions:**

1) What suggestions do you have for the wand?
2) What suggestion do you have for the robot?
3) What suggestions do you have for the screen?